\documentclass[]{aastex701}

\def\figref#1{Figure~\ref{#1}} 
\def\tabref#1{Table~\ref{#1}} 
\def\eqref#1{Equation~\ref{#1}} 
\def\secref#1{Section~\ref{#1}} 

\def\j1641{HESS~J1641$-$463}

\def\chandra{Chandra}
\def\xmm{XMM-Newton}

\def\uG{\hbox{$\mu \mathrm{G}$}}

\def\kms{\hbox{${\rm km}~ {\rm s}^{-1}$}}

\def\Msun{\hbox{$M_{\odot}$}}

\def\vapp{$v_{\rm app}$}
\def\vush{$v^{\prime}_{\rm u}$}
\def\vsh{$v_{\rm sh}$}
\def\vu{$v_{\rm u}$}
\def\vd{$v_{\rm d}$}

\usepackage{acro}

\DeclareAcronym{gde}{
  short = GDE ,
  long  = Galactic Diffuse Emission,
}

\DeclareAcronym{cgb}{
  short = CGB ,
  long  = cosmic gamma-ray background,
}

\DeclareAcronym{grb}{
  short = GRB ,
  long  = gamma-ray burst,
}
 
\DeclareAcronym{agn}{
  short = AGN ,
  long  = active galactic nucleus ,
  long-plural-form = active galactic nuclei
}
\DeclareAcronym{fsrq}{
  short = FSRQ ,
  long  = flat-spectrum radio quasar,
}
  
\DeclareAcronym{lmxb}{
  short = LMXB ,
  long  = low mass X-ray binary ,
  long-plural-form = low mass X-ray binaries ,
}
\DeclareAcronym{hmxb}{
  short = HMXB ,
  long  = high mass X-ray binary ,
  long-plural-form = high mass X-ray binaries ,
}
  
\DeclareAcronym{ic}{
  short = IC ,
  long  = inverse Compton ,
}

\DeclareAcronym{nustar}{
  short = {\it NuSTAR} ,
  long  = Nuclear Spectroscopic Telescope Array ,
}

\DeclareAcronym{bi}{
  short = BI ,
  long  = backside illumination ,
}
\DeclareAcronym{fi}{
  short = FI ,
  long  = frontside illumination ,
}
\DeclareAcronym{fov}{
  short = FoV ,
  long  = field of view ,
}
  
\DeclareAcronym{sim}{
  short = SIM ,
  long  = Science Instrument Module ,
}
\DeclareAcronym{hetg}{
  short = HETG ,
  long  = High Energy Transmission Grating ,
}
\DeclareAcronym{letg}{
  short = LETG ,
  long  = Low Energy Transmission Grating ,
}
\DeclareAcronym{hrc}{
  short = HRC ,
  long  = High Resolution Camera ,
}
\DeclareAcronym{acis}{
  short = ACIS ,
  long  = Advanced CCD Imaging Spectrometer ,
}
\DeclareAcronym{hrma}{
  short = HRMA ,
  long  = High Resolution Mirror Assembly,
}

\DeclareAcronym{compton}{
  short = Compton ,
  long  = Compton Gamma Ray Observatory,
}
\DeclareAcronym{hst}{
  short = HST ,
  long  = Hubble Space Telescope ,
}
\DeclareAcronym{iact}{
  short = IACT ,
  long  = Imaging Atmospheric Cherenkov Telescope ,
}
\DeclareAcronym{wcd}{
  short = WCD ,
  long  = Water Cherenkov Detector ,
}

\DeclareAcronym{hawc}{
  short = HAWC ,
  long  = High-Altitude Water Cherenkov ,
}
\DeclareAcronym{cta}{
  short = CTA ,
  long  = Cherenkov Telescope Array ,
}

\DeclareAcronym{em}{
  short = EM ,
  long  = electromagnetic ,
}
\DeclareAcronym{ism}{
  short = ISM ,
  long  = interstellar medium ,
}
\DeclareAcronym{csm}{
  short = CSM ,
  long  = circumstellar medium ,
}
\DeclareAcronym{sne}{
  short = SNe ,
  long  = supernovae , 
}

\DeclareAcronym{iss}{
  short = ISS ,
  long  = International Space Station ,
}

\DeclareAcronym{uhecr}{
  short = UHECRs ,
  long  = ultra high energy cosmic rays , 
}
\DeclareAcronym{ta}{
  short = TA ,
  long  = Telescope Array , 
}
\DeclareAcronym{auger}{
  short = Auger ,
  long  = Pierre Auger Observatory , 
}
\DeclareAcronym{ams}{
  short = AMS ,
  long  = Alpha Magnetic Spectrometer , 
}
\DeclareAcronym{pamela}{
  short = PAMELA ,
  long  = Payload for Antimatter Matter Exploration and Light-nuclei Astrophysics , 
}
   
\DeclareAcronym{cmb}{
  short = CMB ,
  long  = Cosmic Microwave Background , 
}
\DeclareAcronym{sed}{
  short = SED ,
  long  = spectral energy distribution , 
}
\DeclareAcronym{mhd}{
  short = MHD ,
  long  = magnetohydrodynamical ,
}
\DeclareAcronym{dof}{
  short = dof ,
  long  = degree of freedom ,
}
\DeclareAcronym{cco}{
  short = CCO ,
  long  = central compact object ,
  first-style = default
}
\DeclareAcronym{lmc}{
  short = LMC ,
  long  = Large Magellanic Cloud ,
}
\DeclareAcronym{smc}{
  short = SMC ,
  long  = Small Magellanic Cloud ,
}

\DeclareAcronym{hess}{
  short = H.E.S.S. ,
  long  = High Energy Stereoscopic System ,
  first-style = default
}
\DeclareAcronym{snr}{
  short = SNR ,
  long  = supernova remnant ,
}
\DeclareAcronym{pwn}{
  short = PWN ,
  short-plural = e ,
  long  = pulsar wind nebula ,
  long-plural  = e ,
}

\DeclareAcronym{sn}{
  short = SN ,
  short-plural = e ,
  long  = supernova ,
  long-plural  = e ,
  first-style = default
}

\DeclareAcronym{nw}{
  short = NW ,
  long  = northwest ,
  first-style = default
}
\DeclareAcronym{hxc}{
  short = HXC ,
  long  = hard X-ray component ,
  first-style = default
}

\DeclareAcronym{cr}{
  short = CR ,
  long  = cosmic ray ,
}
\DeclareAcronym{psf}{
  short = PSF ,
  long  = point spread function ,
}
\DeclareAcronym{hpd}{
  short = HPD ,
  long  = half power diameter ,
}
\DeclareAcronym{fwhm}{
  short = FWHM ,
  long  = full width of half maximum ,
}

\DeclareAcronym{pic}{
  short = PIC ,
  long  = particle-in-cell ,
  tag = numerical ,
}
\DeclareAcronym{cxb}{
  short = CXB ,
  long  = Cosmic X-ray Background ,
}
\DeclareAcronym{grxe}{
  short = GRXE ,
  long  = Galactic Ridge X-ray Emission ,
}
\DeclareAcronym{pa}{
  short = PA ,
  long  = Positional Angle ,
}
\DeclareAcronym{dsa}{
  short = DSA ,
  long  = diffusive shock acceleration ,
}

\def\arcmin{\hbox{$^\prime$}}

\def\utw{\smash{\rlap{\lower5pt\hbox{$\sim$}}}}
\def\udtw{\smash{\rlap{\lower6pt\hbox{$\approx$}}}}
\def\apjl{ApJ}%
\def\apjs{ApJS}%
\def\ao{Appl.~Opt.}%
\def\aap{A\&A}%
\usepackage{multirow}
\usepackage{rotating}

\acsetup{patch / longtable = false } 

\def\mycomment#1{\textcolor{cyan}{[NT] #1}} %

\graphicspath{{./}{figures/}}

\begin{document}

\title{
The first proper motion measurement of the acceleration regions in the large-scale jets of SS 433 powering the W50 nebula
}

\author[0000-0001-7209-9204]{Naomi Tsuji}
\affiliation{Institute for Cosmic Ray Research, University of Tokyo, 5-1-5, Kashiwa-no-ha, Kashiwa, Chiba 277-8582, Japan}
\affiliation{Faculty of Science, Kanagawa University, 3-27-1 Rokukakubashi, Kanagawa-ku, Yokohama, Kanagawa 221-8686}
\affiliation{Interdisciplinary Theoretical \& Mathematical Science Center (iTHEMS), RIKEN, 2-1, Hirosawa, Wako, Saitama 351-0198, Japan}
\affiliation{Department of Physics, Rikkyo University, Nishi-Ikebukuro 3-34-1, Toshima-ku, Tokyo, 171-8501, Japan}
\email[show]{ntsuji@icrr.u-tokyo.ac.jp} 
\author[0000-0002-7272-1136]{Yoshiyuki Inoue}
\affiliation{Department of Earth and Space Science, Graduate School of Science, Osaka University, Toyonaka, Osaka 560-0043, Japan}
\affiliation{Interdisciplinary Theoretical \& Mathematical Science Center (iTHEMS), RIKEN, 2-1, Hirosawa, Wako, Saitama 351-0198, Japan}
\affiliation{Kavli Institute for the Physics and Mathematics of the Universe (WPI), The University of Tokyo, Kashiwa 277-8583, Japan}
\email{yinoue@astro-osaka.jp}
\author[0000-0002-7576-7869]{Dmitry Khangulyan}
\affiliation{Key Laboratory of Particle Astrophyics, Institute of High Energy Physics, Chinese Academy of Sciences, 100049 Beijing, China}
\affiliation{Tianfu Cosmic Ray Research Center, 610000 Chengdu, Sichuan, China}
\email{khangulyan@ihep.ac.cn}
\author[0000-0002-9709-5389]{Kaya Mori}
\affiliation{Columbia Astrophysics Laboratory, Columbia University, 538 West 120th Street, New York, NY 10027, USA}
\email{kaya@astro.columbia.edu}
\author[0000-0001-6189-7665]{Samar Safi-Harb}
\affiliation{Department of Physics and Astronomy, University of Manitoba, Winnipeg, MB R3T 2N2, Canada}
\email{Samar.Safi-Harb@umanitoba.ca}
\author[0000-0002-4383-0368]{Takaaki Tanaka}
\affiliation{Konan University, Department of Physics, 8-9-1 Okamoto, Higashinada, Kobe, Hyogo, Japan, 658-8501}
\email{ttanaka@konan-u.ac.jp}
\author[0000-0002-9105-0518]{Laura Olivera-Nieto}
\affiliation{Max-Planck-Institut f\"{u}r Kernphysik, Saupfercheckweg 1, 69117 Heidelberg, Germany}
\email{laura.olivera-nieto@mpi-hd.mpg.de}
\author[0000-0003-0146-3691]{Brydyn Mac Intyre}
\affiliation{Department of Physics and Astronomy, University of Manitoba, Winnipeg, MB R3T 2N2, Canada}
\email{macintyb@myumanitoba.ca}
\author[0000-0002-3562-6965]{Kazuho Kayama}
\affiliation{Department of Physics, Graduate School of Science, Kyoto University, Kitashirakawa Oiwake-cho, Sakyo-ku, Kyoto 606-8502, Japan}
\affiliation{Remote Sensing Technology Center of Japan, 3-17-1, Toranomon, Minato-ku, Tokyo, 105-0001, Japan}
\email{kayama.kazuho.57r@kyoto-u.jp}
%
%
\author[0000-0002-5504-4903]{Takeshi Go Tsuru}
\affiliation{Department of Physics, Graduate School of Science, Kyoto University, Kitashirakawa Oiwake-cho, Sakyo-ku, Kyoto 606-8502, Japan}
\email{tsuru@cr.scphys.kyoto-u.ac.jp}
\author[0000-0003-1518-2188]{Hiroyuki Uchida}
\affiliation{Department of Physics, Graduate School of Science, Kyoto University, Kitashirakawa Oiwake-cho, Sakyo-ku, Kyoto 606-8502, Japan}
\email{uchida@cr.scphys.kyoto-u.ac.jp}
\author[0009-0002-9370-3313]{Tatsuki Fujiwara}
\affiliation{Department of Earth and Space Science, Graduate School of Science, Osaka University, Toyonaka, Osaka 560-0043, Japan}
\email{u000719k@ecs.osaka-u.ac.jp}
\author[0000-0003-1157-3915]{Felix Aharonian}
\affiliation{Yerevan State University, 1 Alek Manukyan Street, Yerevan 0025, Armenia}
\affiliation{University of Science and Technology of China, 230026 Hefei, Anhui, China}
\affiliation{Max-Planck-Institut f\"{u}r Kernphysik, Saupfercheckweg 1, 69117 Heidelberg, Germany}
\affiliation{Dublin Institute for Advanced Studies, School of Cosmic Physics, 31 Fitzwilliam Place, Dublin 2, Ireland}
\email{felix.aharonian@mpi-hd.mpg.de}

\begin{abstract}

We report on new Chandra ACIS-I observations of the X-ray knots located in the western and eastern lobes of W50 associated with the parsec-scale jets of the Galactic microquasar SS 433. These knots are likely counterparts of the recently detected very-high-energy ($E>100$~GeV) gamma-ray emission by HAWC and H.E.S.S. These findings, together with the ultra-high-energy signal recently reported by the LHAASO collaboration, have established the SS~433/W50 system as a unique jet-driven PeVatron candidate. Combining new and archival Chandra data, we perform the first proper motion search of the X-ray knot structures over a time interval spanning approximately 20 years. We found no statistically significant motion of these knots at the 3$\sigma$ confidence level, and place an upper limit of $<$ 0.019--0.033$c$ (5,800--9,800 \kms) for the speed of the innermost knots at an assumed distance $d=5.5$ kpc. Combined with the velocities reported in the literature, the upstream speed in the shock rest frame would reach several 10$^4$ \kms, suggesting that highly efficient particle acceleration, approaching the Bohm limit, is occurring. The absence of significant motion of the knots suggests the presence of a standing recollimation shock, formed by the balance between the jet pressure and the external pressure.  This interpretation is consistent with the expected occurrence of such shocks at 20--30 pc from SS~433, matching the location of the observed knots.


\end{abstract}



\section{Introduction} \label{sec:intro}



Microquasars --- X-ray binaries with jets --- have emerged as a new class of sources in the very-high-energy (VHE; $E> 100$ GeV) and ultra-high-energy (UHE; $E>100$ TeV) bands.
To date, VHE-UHE emission was detected from five sources:
SS 433 \citep{abeysekara_very-high-energy_2018,hess_2024_ss43}, V4641 Sgr \citep{alfaro_ultra-high-energy_2024}, GRS~1915$+$105, MAXI J1820$+$070, and Cyg~X-1 \citep{lhaaso_2024_microquasars}.
The gamma-ray spectra of these objects are extended up to the $\lesssim$ PeV energy range, suggesting that the parent particles are accelerated up to, and even beyond, PeV energies.
This establishes microquasars as a new Galactic class of PeV particle accelerators (PeVatrons) or even super-PeVatrons (i.e., sources capable of boosting particle energy  beyond the PeV limit).
However, it remains unclear which particles (electrons or protons) are responsible for the gamma-ray emission, where within microquasars particle acceleration occurs, and what physical mechanisms drive this acceleration.


The microquasar SS 433 and its nebula W50 represent the first microquasar detected in the VHE band \citep{abeysekara_very-high-energy_2018}.
The central engine consists of a $\sim$5--15 \Msun\ black hole and a $\sim10$ \Msun\ type A supergiant star orbiting each other with a period of 13.1 days \citep{kotani_iron-line_1996,2004ASPRv..12....1F,hillwig_spectroscopic_2008,kubota_subaru_2010,seifina_nature_2010,bowler_ss_2018,cherepashchuk_mass_2019}.
Bipolar jets emerging from the compact object have been identified in radio observations \citep[e.g., ][]{blundell_symmetry_2004,blundell_ss433s_2018,marti_radio_2018}.

Initial X-ray surveys with ROSAT, ASCA, and RXTE
revealed X-ray knot-like structures aligned along both the eastern and western jet axes, extending tens of parsecs from SS~433 and dubbed as e1, e2, e3 (in the eastern lobe) and w1, w2 (in the western lobe), 
as shown in \figref{fig:image} 
\citep{safi-harb_rosat_1997,safi-harb_rossi_1999}.
The X-ray emission in the e3 knot is dominated by soft thermal X-ray emission and is
located at the eastern edge of W50, coinciding with the so-called radio ear, which is likely a termination shock of the system.
Notably, the spectra were found to soften gradually away from SS~433 and are characterized by a hard non-thermal spectrum in e1--e2 and w1--w2 \citep{safi-harb_rosat_1997,safi-harb_rossi_1999,brinkmann_xmm-newton_2007} leading to the first suggestion that W50 represents an excellent Galactic particle accelerator up to at least a few 100 TeV energies \citep{safi-harb_rossi_1999, 1998NewAR..42..579A}. This motivated the search for TeV gamma-rays from the SS~433/W50 system \citep{2018A&A...612A..14M}.

More recent X-ray observations acquired with XMM-Newton and NuSTAR have shown that the non-thermal X-ray emission in the eastern lobe starts closer to SS~433 than the previously defined e1 region \citep{safi-harb_hard_2022}.
We refer to this newly identified region as e0 or head hereafter\footnote{The e1 region was initially defined conservatively, as it was unclear in the 1990s whether the e0/head region was extended or a point source.}.
Recently, VHE gamma rays were detected with the High-Altitude Water Cherenkov Observatory (HAWC) at the inner X-ray knots e0--e2 and w1--w2 \citep{abeysekara_very-high-energy_2018}. 
Furthermore, observations with the better angular resolution of the High Energy Stereoscopic System (H.E.S.S.) unveiled that the higher-energy gamma-ray emission ($E>10$ TeV) originates from regions close to the innermost knots, e0/head and w1, confirming these shock regions as acceleration sites \citep{hess_2024_ss43}. 
Motivated by these gamma-ray findings, an extensive follow-up X-ray observation campaign is currently underway \citep{safi-harb_hard_2022,chi_x-ray_2024,kaaret_x-ray_2024,macIntyre2025}. 

The origin of gamma rays --- whether leptonic or hadronic --- has been controversial in the SS 433/W50 system.
The similar morphology observed in non-thermal X-rays and TeV gamma-rays suggests a leptonic origin.
It has been proposed that accelerated electrons are produced at the innermost knots, e0/head and w1, which have the hardest X-ray spectra with a photon index $\Gamma \approx 1.6$ and the higher energy gamma-ray emission. These electrons are then advected and cooled downstream toward the outer lobes. This model is well reconciled with the observed gradual softening of the X-ray and gamma-ray emission \citep[e.g., ][]{sudoh_multiwavelength_2020,kimura_deciphering_2020,kayama_spatially_2022,hess_2024_ss43}.
An abrupt brightening at the outer X-ray knots, e2 and w2, is likely due to magnetic field amplification \citep{kayama_spatially_2022,kayama_x-ray_2025}.
More recently, LHAASO detected UHE emission, not just from the eastern and western lobes of W50, 
but also remarkably even higher energy photons from the northern part of the W50 nebula which is spatially coincident with a dense HI cloud \citep{lhaaso_2024_cygnus,su_large-scale_2018}.
This northern UHE emission might be hadronic because of the suppression of the inverse Compton (IC) emission in the UHE range, potentially providing the first evidence of the SS~433/W50 system as a hadronic PeVatron.

Although the aforementioned findings indicate the existence of accelerated electrons at the X-ray knots and energetic protons at the northern region,
the exact mechanism of particle acceleration remains poorly understood.
Deciphering the dynamics in the SS 433/W50 system is crucial for understanding the acceleration mechanism.
There are different observational approaches to constrain velocities of astrophysical outflow;
(1) measuring a shift of line emission that constrains a line-of-sight speed and
(2) measuring a shift of spatial structure that constrains a proper motion speed.
The speed of the jet from the microquasar was measured to be $\sim 0.26c$ by measuring the Doppler shift of line emissions and taking into account the jet precession period and geometry \citep{margon_ten_1989,marshall_high-resolution_2002,marshall_multiwavelength_2013}. 
\cite{sakemi_energy_2021} placed an upper limit of $0.023c$ at the radio ear, e3, based on non-detection of proper motions by the VLA data over a time interval of 33 years.
The downstream velocity at locations of the w1/e0 knots was constrained to be
0.045--0.08$c$ by modeling the gamma-ray observations \citep{hess_2024_ss43} and 
0.065--0.1$c$ by the X-ray observations \citep{kayama_spatially_2022,kayama_x-ray_2025}.
However, no direct measurement of the dynamics at the inner X-ray knots has been made to date.

In this paper, we present the first proper motion measurements of the X-ray knots
by utilizing new and archival \chandra\ data, which offer the best angular resolution among X-ray satellites, over a time interval spanning $\sim$20 years.
This close-up view will have implications on probing the acceleration mechanism and efficiency in the acceleration sites, as well as the jet dynamics in the SS~433/W50 system. 
A detailed spectral analysis and spectral evolution over 20 years will be presented in a separate publication.
The dataset and data reduction are given in \secref{sec:observations}, and the analysis and results are detailed in 
\secref{sec:analysis}.
Finally, we present the discussion of our results and implication on particle acceleration in Section \ref{sec:discussion}.

\section{Observations and data reduction}
\label{sec:observations}

We have performed \chandra\ observations of the western knots (w1 and w2, each with an exposure time of 40 ks) in 2023
and the eastern knots (e0/head with 70 ks and e2 with 50 ks) in 2024, as summarized in Appendix (\tabref{tab:dataset}).
In this paper, we also made use of the archival \chandra\ and \xmm\ data (\tabref{tab:dataset}).
It should be noted that there are no on-axis archival observations of the w1 and e0 knots.
However, the two archival \chandra\ data (ObsIDs 3790 and 3843) partially covered w1 and e0, and thus would be useful for comparison with the new observations.

The \chandra\ data were reprocessed by using {\tt chandra\_repro} with CALDB version 4.9.6 in CIAO version 4.14, the analysis software provided by the Chandra X-Ray Center\footnote{\url{https://cxc.harvard.edu/}}.
We analyzed the \xmm\ data with the XMM-Newton Science Analysis System (SAS, version 20.0). 
The \xmm\ data were reprocessed by {\tt emchain} and {\tt epchain}, 
and the contamination of flares was filtered by {\tt mos-filter} and {\tt pn-filter}.

\section{Analysis and Results}
\label{sec:analysis}


\begin{sidewaysfigure}[h!]
\centering
\plotone{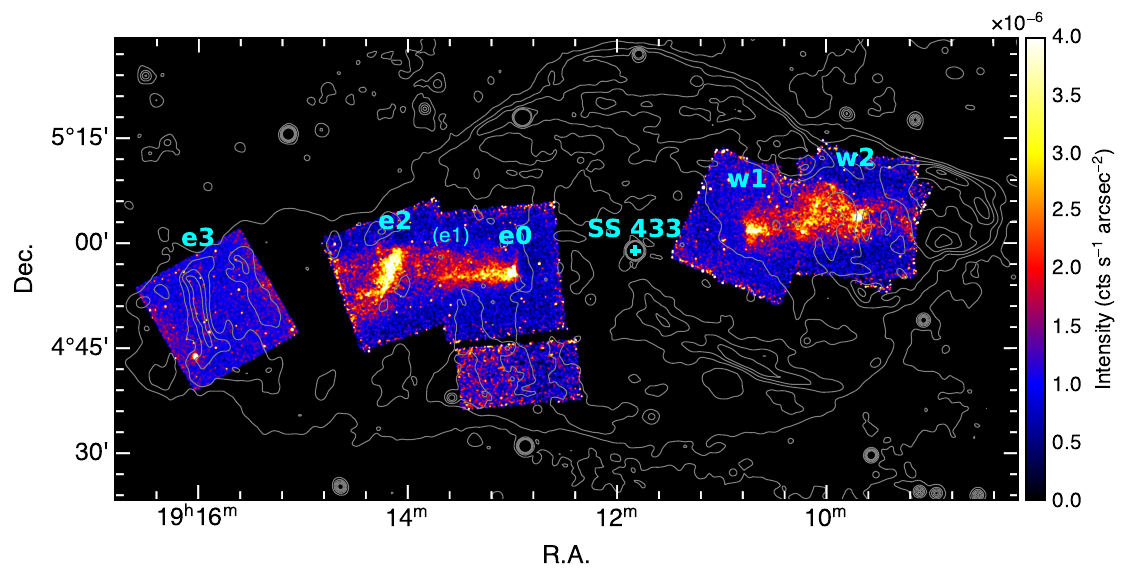}
\includegraphics[width=\linewidth]{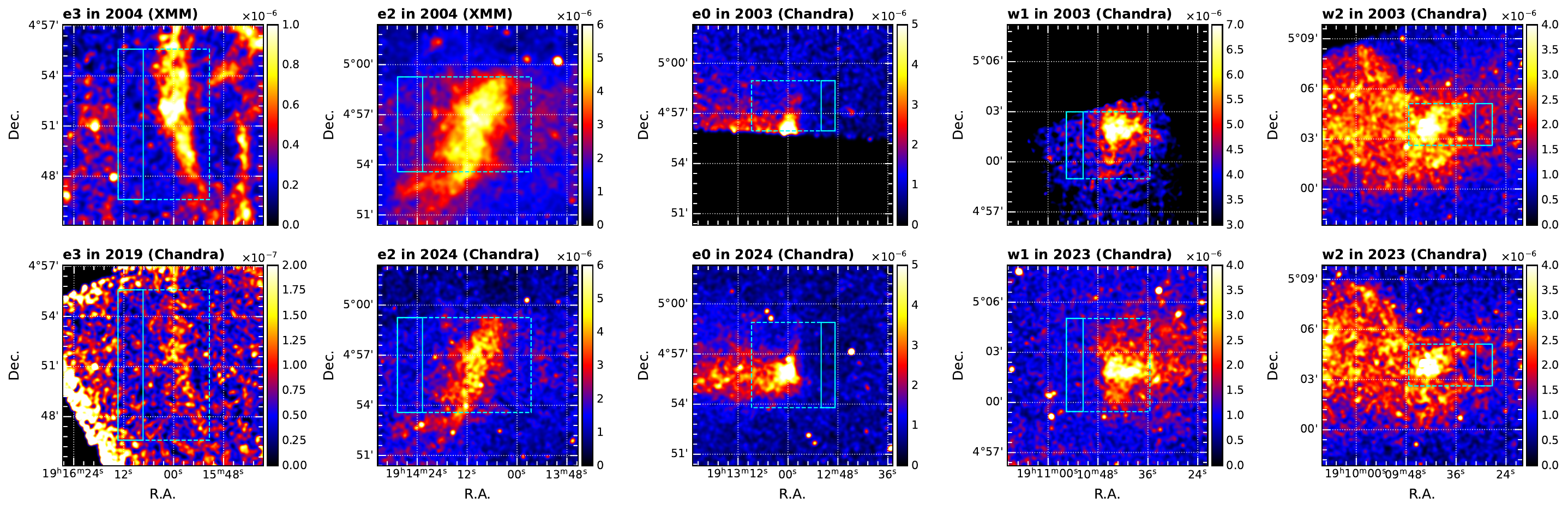}
\caption{ 
Top: The new 0.5--7 keV \chandra\ image taken in 2023--2024 (e3 in 2019), with a spatial bin size of 2 arcsec.
The white contours indicate the radio continuum radiation at 1.4 GHz \citep{dubner_high-resolution_1998}.
Bottom: zoomed-in images of the X-ray knots and comparison of the archival and new data.
The cyan dashed box indicates the region used to produce projection profiles in \figref{fig:projection}, while the solid box shows the background region.
The energy band is 0.5--7 keV, except 0.5--1.2 keV for the e3 knot.
The pixel sizes of the \chandra\ and \xmm\ images are 2 arcsec and 2.5 arcsec, respectively.
The maps are shown in units of counts~s$^{-1}$~arcsec$^{-2}$.
\label{fig:image}}
\end{sidewaysfigure}


We combined all epochs of the newly observed \chandra\ data by using {\tt merge\_obs} and produced the flux images shown in \figref{fig:image}.
The archival \chandra\ data images were generated in the same way using {\tt fluximage}.
For \xmm, exposure-corrected images were produced using a standard procedure, as summarized in the SAS Thread\footnote{\url{https://www.cosmos.esa.int/web/xmm-newton/sas-thread-esasimage}}
\cite[e.g., ][]{chi_x-ray_2024}.

\begin{figure}[ht!]
\centering
\includegraphics[width=0.7\linewidth]{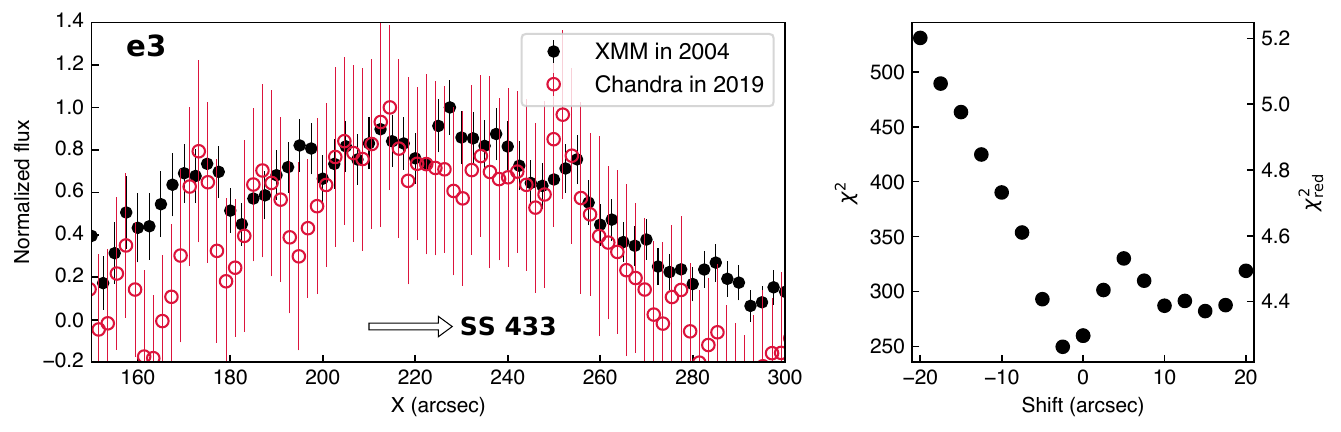} \\
\includegraphics[width=0.7\linewidth]{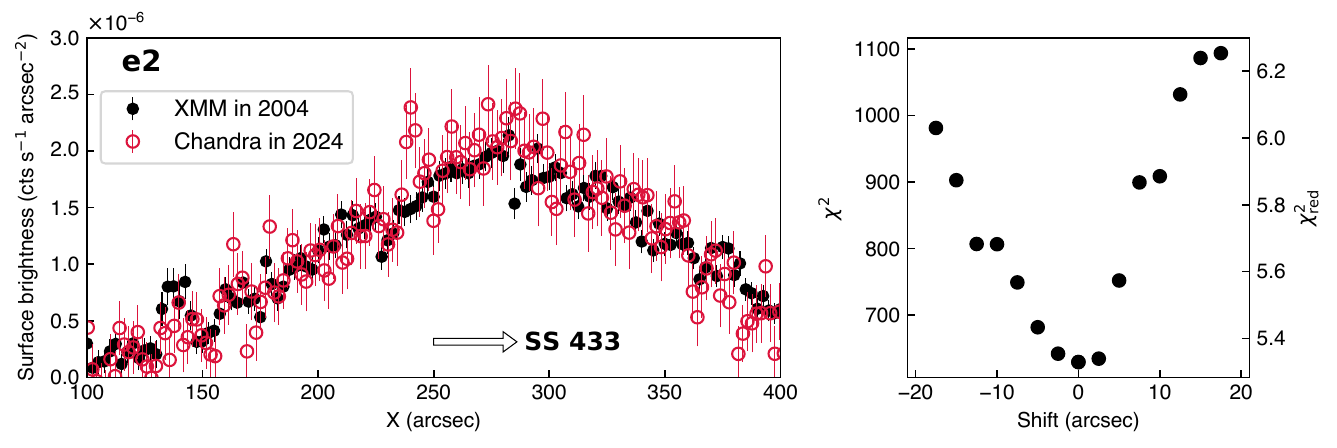} \\
\includegraphics[width=0.7\linewidth]{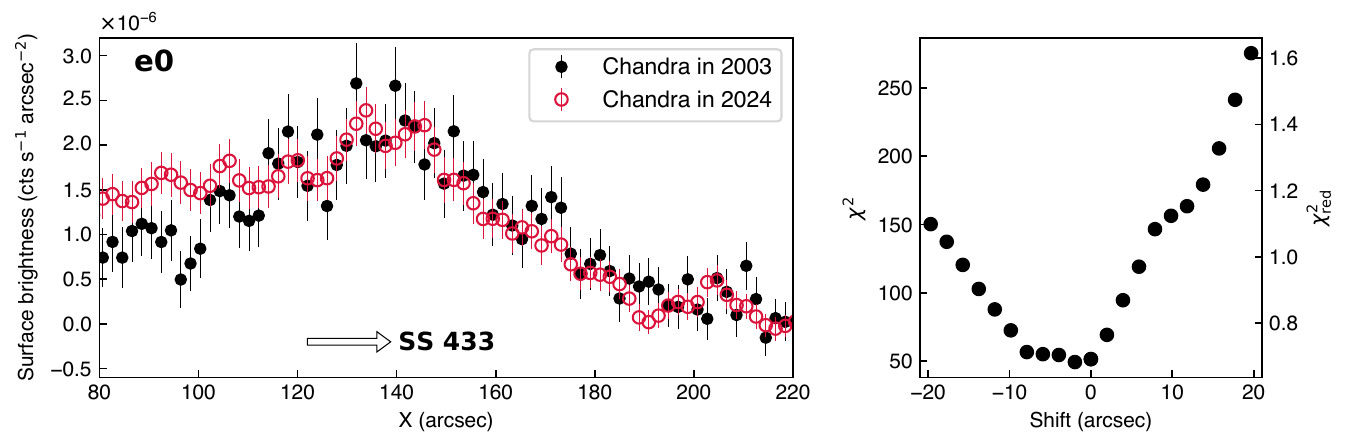} \\
\includegraphics[width=0.7\linewidth]{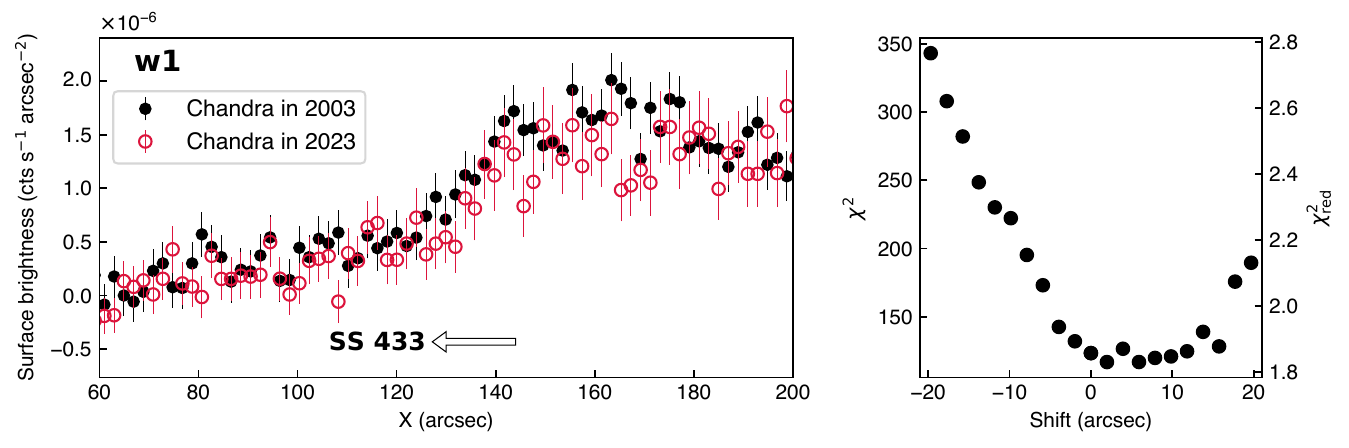} \\
\includegraphics[width=0.7\linewidth]{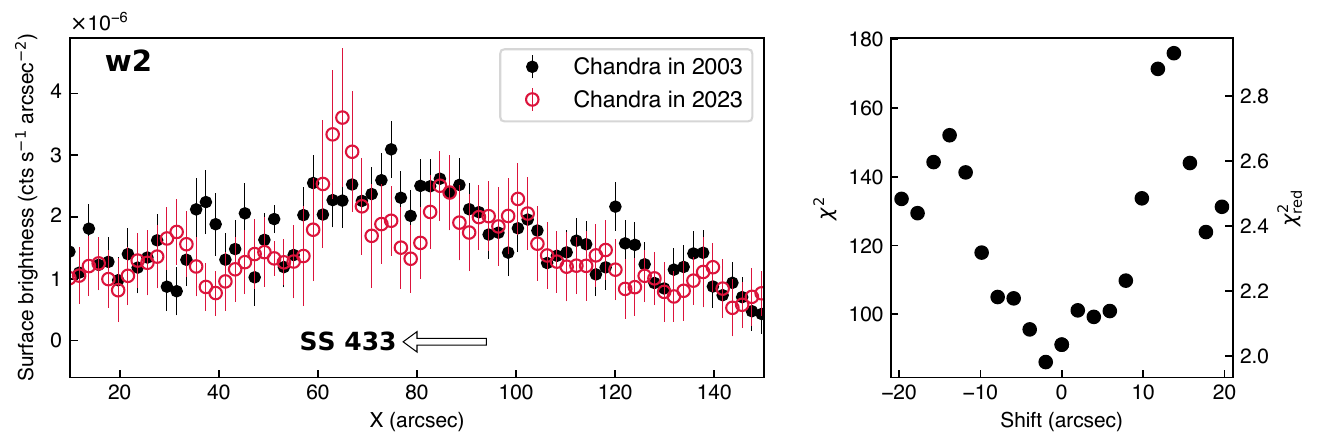} \\
\caption{
Left: 
The projection profiles, extracted from the regions shown in \figref{fig:image}. 
Right: the $\chi^2$ profiles, where the positive shift indicates a direction to the west. 
\label{fig:projection}
}
\end{figure}

As shown in \figref{fig:image}, we defined a box region around each knot and made a projection profile along the right ascension by integrating the pixel values along the declination.
First, the background level was estimated from a source-free region and subtracted from the projection profile.
Next, the profile was normalized by dividing it by the area of the projection slice (i.e., the spatial grid times the height of the box) at each bin to enable comparison across different epochs, as the knot region was not fully covered in the archival data.
For e3, the profile was normalized by its maximum value, because we compared the \chandra\ and \xmm\ images in the soft (0.5--1.2 keV) X-ray band, where the effective area of \chandra\ is small and the absorption becomes large.
We checked the results did not depend on the normalization method. 
It should be noted that the flux time variation within the box is less than 30\%, which does not significantly affect the following results. 
\figref{fig:projection} shows the obtained projection profiles of e3, e2, e0, w1, and w2.


The proper motion of each knot was measured as described below \citep[see also, e.g., ][]{Katsuda2010_tycho,tsuji_expansion_2016,tanaka_shock-cloud_2020}.
From the projection profile, we calculated 
\begin{eqnarray}
    \chi^2 = \sum_i \frac{ \left(f_i - m_i\right)^2}{ \sigma_i^2  }, 
    \label{eq:chi2}
\end{eqnarray}
where 
$i$ is each bin in the projection profile, 
$f_i$ is the profile of the data (the archival data were adopted, unless otherwise mentioned),
$m_i$ is the profile of the model (the new \chandra\ data),
and $\sigma_i$ indicates the error of $f_i$.
The $\chi^2$ curve was produced by shifting the profile of the archival data within the range of $-$20 arcsec to $+$20 arcsec and calculating the $\chi^2$ value at each shift (\figref{fig:projection}).
The best-fit shift is given by $\chi^2 = \chi^2_{\rm min}$, and
90\% uncertainty can be derived by $\chi^2 = \chi^2_{\rm min} + 2.7 $.
To obtain the $\chi^2_{\rm min}$ value and uncertainty, we conducted two methods:
(1) modeling the $\chi^2$ curve with a quadratic function, and
(2) interpolating the $m_i$ data in \eqref{eq:chi2} to calculate the $\chi^2$ curve in detail.
We confirmed that these methods did not produce significant differences and adopted method (1) in the following.
We measure an angular shift of
1.6 $\pm$ 1.0  arcsec in e3, 0.9 $\pm$ 1.0  arcsec in e2, 2.1 $\pm$ 1.4  arcsec in e0/head, 3.4 $\pm$ 2.3  arcsec in w1, $-$1.1 $\pm$ 1.6  arcsec in w2,
where positive values indicate a direction toward the outer region (i.e., the positive shift in e0--e3 corresponds to motion to the east, while that in w1--w2 means motion to the west).
We do not detect significant shifts in any of the knots at 3$\sigma$ confidence level (C.L.), and we therefore provide the 3$\sigma$ uncertainty as well (\tabref{tab:speed}).

Next, we take the systematic uncertainty into consideration.
First, we evaluate the pointing accuracy by measuring the astrometry of point-like sources in the field of view (FoV) of the new and archival data, resulting in an accuracy of 0.88 arcsec\ in average.
It is worth noting that the performance of \chandra\ has degraded between 2003 and 2023--2024.
For example, the effective area has declined significantly in the low energy range of $<2$~keV due to the accumulation of molecular contamination \citep[e.g., ][]{grant_advanced_2024}\footnote{See also \url{https://cxc.harvard.edu/proposer/POG/html/chap6.html}}.
However, the degrading is not an issue in the hard X-ray band ($>2$ keV), and thus it does not affect the flux of the X-ray knots in SS~433 (except for e3) being dominated by non-thermal hard X-ray emission.
We also find that the result of the best-fit shift value does not largely depend (by less than 30\%) on the following setup:
flipping the new and archival data in \eqref{eq:chi2},
the extraction region,  
the angle of the region,
the smoothing radius,
the spatial bin,
and the energy bands.
%
The half-power diameter of the point spread function (PSF) of the new \chandra\ data is 0.5 arcsec, 
while that of the archival \chandra\ data is $\sim$10 arcsec because of its 14--16 arcmin off-axis pointing and that of the \xmm\ data is 15 arcsec.
However, the PSF value does not affect the proper motion measurement since we have enough photon statistics, the pointing accuracy is much smaller than the PSF of the archival data, and the knot structure is more extended than the PSF.

\begin{figure}[ht!]
\plotone{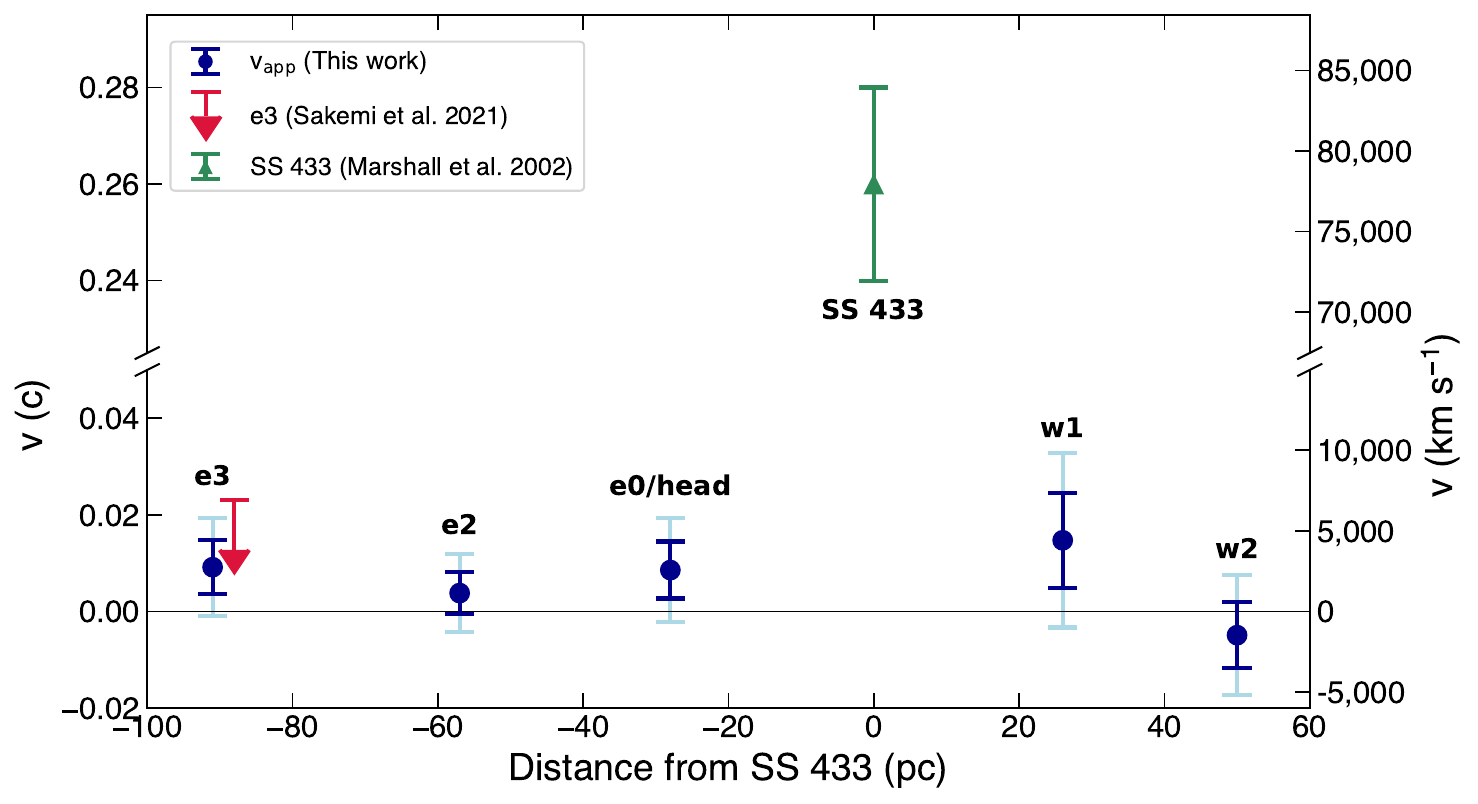}
\caption{
The velocity profile in the SS~433/W50 system, assuming $d=5.5$ kpc.
The blue and lightblue errorbars show the 90\% and 3$\sigma$ uncertainties, respectively.
The positive velocity is defined as the direction toward the outer lobe away from SS~433 (i.e., the positive values of the eastern knots mean motions to the east, while the positive values of the western knots mean motions to the west).
The velocities of the knots (e0--e3 and w1--w2) were measured by proper motion studies (this work, and \cite{sakemi_energy_2021} for e3), while that of SS~433 was derived by the shift of line emission in the arcsec-scale X-ray jets \citep{marshall_high-resolution_2002}.
\label{fig:speed}}
\end{figure}

\tabref{tab:speed} summarizes the proper motions, defined as the measured shift divided by the time interval between the new and archival data spanning $\sim$20 years.
We compute the velocity assuming a distance of $d=5.5$~kpc \citep{hjellming_analysis_1981}.
It should be noted that there is relatively large uncertainty on the estimated distance;
$\sim$5.5 kpc by VLA observations and modeling \citep{hjellming_analysis_1981}, $\sim$3 kpc by HI gas \citep{dubner_high-resolution_1998}, 4.5 $\pm$ 0.2 kpc by VLBI \citep[e.g., ][]{marshall_multiwavelength_2013}, and $\sim$3.8 kpc by the parallax measurement with Gaia \citep{arnason_distances_2021}.
\figref{fig:speed} illustrates the velocity profile in the SS~433/W50 system, combined with the speed of SS~433 and the e3 knot in the literature \citep{marshall_high-resolution_2002,marshall_multiwavelength_2013,sakemi_energy_2021}.
The obtained apparent speed ($v_{\rm app}$) at the assumed distance of 5.5~kpc is (0.0038--0.15)$c$, and the 3$\sigma$ upper limit is (0.012--0.033)$c$.
We found that the knots exhibit clear deceleration compared to the base jet of SS~433 ($v_{\rm jet} \sim 0.26c$).
Note that the PSF of the archival data, $\sim$10 arcsec, corresponds to $v_{\rm app} \sim 0.04c$.

\begin{table}[ht!]
\begin{center}
\caption{Proper motion of the X-ray knots}
\begin{tabular}{ c | ccc | ccc | ccc }
\hline \hline 
   \multirow{2}{*}{Knot} & \multicolumn{3}{c | }{Motion  (arcsec~yr$^{-1}$)} & \multicolumn{3}{c |}{$v_{\rm app}$ (c) }  & \multicolumn{3}{c}{$v_{\rm app}$ (\kms) }   \\ 
    & Best fit & 90\% C.L. & 3$\sigma$ C.L. &  Best fit &  90\% C.L.  & 3$\sigma$ C.L. &   Best fit & 90\% C.L.  & 3$\sigma$ C.L.  \\
    \hline
e3 & 0.11  & $\pm$ 0.06 & $\pm$ 0.12   &   0.0092 & $\pm$ 0.0056 & $\pm$ 0.010   &   2,700 & $\pm$ 1,700 & $\pm$ 3,000  \\ 
e2 & 0.044  & $\pm$ 0.050 & $\pm$ 0.092   &   0.0038 & $\pm$ 0.0044 & $\pm$ 0.008   &   1,100 & $\pm$ 1,300 & $\pm$ 2,400  \\ 
e0/head & 0.098  & $\pm$ 0.068 & $\pm$ 0.120   &   0.0086 & $\pm$ 0.0059 & $\pm$ 0.011   &   2,600 & $\pm$ 1,800 & $\pm$ 3,200  \\ 
w1 & 0.17  & $\pm$ 0.11 & $\pm$ 0.21   &   0.015 & $\pm$ 0.0099 & $\pm$ 0.018   &   4,400 & $\pm$ 3,000 & $\pm$ 5,400  \\ 
w2 & -0.057  & $\pm$ 0.078 & $\pm$ 0.140   &   -0.0049 & $\pm$ 0.0068 & $\pm$ 0.012   &   -1,500 & $\pm$ 2,000 & $\pm$ 3,700  \\ 
    \hline
    \end{tabular}
    \label{tab:speed}
\end{center}
    \tablecomments{
    $d=5.5$~kpc is assumed.
The positive velocity is defined as the direction toward the outer lobe away from SS~433 (i.e., the positive values of the eastern knots mean motion to the east, while the positive values of the western knots mean motion to the west).
}
\end{table}

\section{Discussion} \label{sec:discussion}


The velocity measured in this paper is an apparent velocity seen by the observer, referred to as $v_{\rm app}$.
In the case of a quasi-stationary outflow, this speed should reflect the shock speed, and thus we assume $v_{\rm sh} = v_{\rm app}$.
For a strong shock in a non-relativistic polytropic gas, the Rankine-Hugoniot condition is described as 
\begin{eqnarray}
    v^{\prime}_{\rm u} = v_{\rm u} - v_{\rm sh} = 4 ( v_{\rm d} - v_{\rm sh} )  = 4 v^\prime_{\rm d},
    \label{eq:RH}
\end{eqnarray}
where \vu\ and \vd\ are the upstream and downstream speeds in the observer's frame, respectively.
The speed in the shock rest frame is denoted with a prime symbol (i.e., $v^\prime$). 
One can derive \vu\ from \eqref{eq:RH} as  
\begin{eqnarray}
    v_{\rm u} = 4 v_{\rm d} - 3 v_{\rm sh} .
    \label{eq:vu}
\end{eqnarray}
Therefore, the measured speed, $v_{\rm sh}$, combined with \vd, can be used to constrain \vu\ and \vush.
The downstream velocity can be constrained by modeling the processes of acceleration and advection occurring in the lobes and comparing the results with X-ray and gamma-ray observations \citep[e.g., ][]{sudoh_multiwavelength_2020}.
For example, \vd\ was obtained to be 0.045--0.083$c$ \citep[0.021--0.12$c$ including the systematic uncertainty;][]{hess_2024_ss43} or 0.065--0.1$c$ \citep{kayama_spatially_2022}.
As the jet speed at the launching site of the microquasar SS~433 is measured as $\sim 0.26 c$ \citep[e.g., ][]{marshall_high-resolution_2002}, we assume $ 0 \leq v_{\rm u} \leq 0.26c$ and $0 \leq v^\prime_{\rm u} \leq 0.26c$. 

\subsection{Particle acceleration}

In the context of particle acceleration at the plausible sites --- the innermost knots (e0/head and w1) --- the shock speed, \vush, plays a crucial role in constraining the acceleration mechanism.
Our measurement reveals that the speed of the e0 knot is \vsh\ = (0.0086 $\pm$ 0.0059)$c$ with the 3$\sigma$ confidential interval (C.I.) of $-$0.002$c$ to $ 0.019c$.
In the \cite{hess_2024_ss43} study, \vd\ was derived as 0.045$c$ in the constant-speed and expanding jet model, resulting in \vush\ of (0.15 $\pm$ 0.02)$c$ corresponding to 44,000 $\pm$ 5,000~\kms\ (0.10--0.19$c$ and 31,000--57,000 \kms\ in 3$\sigma$ C.I.). 
In the constant-speed and non-expanding jet case, \vd\ is 0.061$c$, yielding and \vush\ of (0.21$\pm$0.02)$c$ and 63,000$\pm$5,000 \kms\ (0.17--0.25$c$ and 50,000--76,000~\kms\ in 3$\sigma$ C.I.). 
Even adopting the lower limit of \vd\ $= 0.021c$ \citep[which includes the systematic uncertainty; ][]{hess_2024_ss43}, \vush\ is 0.05$\pm 0.024 c$ and 15,000$\pm$7,000 \kms\ (0.007--0.093$c$ and 2,000--28,000 \kms\ in 3$\sigma$ C.I.).
Similar constraints were obtained for the w1 knot, as summarized in Table~\ref{tab:Vu} and Appendix~B.
Such high values of \vush\ enable efficient particle acceleration through, e.g., diffusive shock acceleration (DSA).
If \vd\ $\gtrsim 0.08 c$, as assumed in the decelerating and expanding jet model of \citet{hess_2024_ss43} and in some cases considered by \citet{kayama_x-ray_2025}, the scenario appears unlikely, since it would imply \vu\ $>0.26c$, exceeding $v_{\rm jet}$.

\begin{figure}[b!]
\plotone{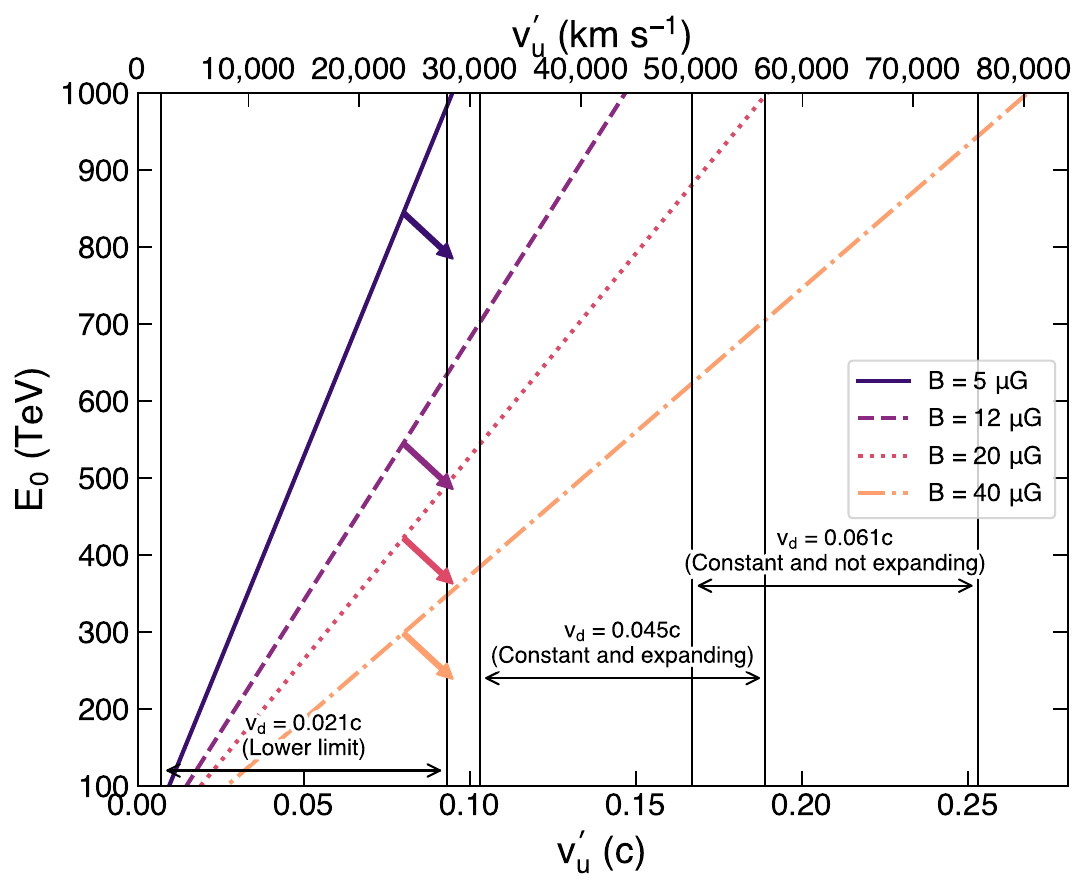}
\caption{
The relation between the shock velocity (\vush) and the electron cutoff energy ($E_0$) in \eqref{eq:eta} with different values of the magnetic field.
The colored diagonal lines indicate the Bohm-limit cases with $B$ of 5, 12, 20, and 40~\uG, and the arrows show the allowed parameter space with $\eta_{\rm B} > 1$.
The vertical lines denote the constrained values of \vush\ for the e0 knot, assuming the constant-speed and expanding jet model (\vd\ = 0.045$c$), the constant-speed and non-expanding jet (\vd\ = 0.061$c$), and the lower limit of \vd\ = 0.021$c$ \citep{hess_2024_ss43}.
\label{fig:eta}}
\end{figure}

In DSA, the acceleration efficiency is sometimes characterized by the so-called Bohm factor, $\eta_\mathrm{B}$ ($>1$), which is defined as the ratio of the particle mean free path to Larmor radius.
In the loss-limited case, equating the timescale of acceleration and that of energy loss mainly due to synchrotron radiation yields the $\eta_{\rm B}$ parameter as 
\begin{eqnarray}
\eta_\mathrm{B} &=& 1.3 \, \left(\frac{ E_0 }{200~{\rm TeV}} \right)^{-2} \left( \frac{ v^\prime_{\rm u} }{10,000\, \mathrm{km\,s}^{-1} } \right)^2
\left( \frac{B}{12\, \mu\mathrm{G}} \right)^{-1} , \label{eq:eta}
 \label{eq:Electron0} 
\end{eqnarray}
where $E_0$ and $B$ are the electron cutoff energy parameter and the strength of magnetic field, respectively.
In the acceleration sites (e0/head and w1), the X-ray spectra are hard with $\Gamma \approx 1.6$ and show no significant sign of cutoff \citep{safi-harb_hard_2022,macIntyre2025}.
By fitting the gamma-ray spectra, the electron cutoff energy was constrained to be $>200$~TeV assuming the electron spectral index, $s_e$, of 2 \citep{hess_2024_ss43} and $\sim 200$~TeV with $s_e = 2.4$ \citep{lhaaso_2024_microquasars}.
By applying the IC model in \cite{Zirakashvili2007} to the gamma-ray spectra of the LHAASO data of W50, we also obtain an $E_0$ of 200 TeV.
However, these gamma-ray spectra are spatially integrated over the entire eastern and western lobes, and therefore would not represent the spectra of the innermost knots.
Determining the maximum energy from the innermost knot regions is beyond the scope of this study, and
we assume a plausible and conservative range of $E_0$ between 100 TeV and 1 PeV.
Another key parameter in determining $\eta_\mathrm{B}$ is the magnetic field strength.
An equipartition magnetic field of $\sim12$~\uG\ was derived at the e0 knot \citep{safi-harb_hard_2022}.
An average magnetic field of $B \sim 20$~\uG\ has been reported for both the eastern and western lobes by fitting the broadband SED in the leptonic model \citep{hess_2024_ss43}.
The detailed modeling of the X-ray spectral profile along the jet suggests that the magnetic field is 5--16 \uG\ within the lobes except the outer knots of e2 and w2 and locally enhanced up to $\sim 40$~\uG\ at e2 and w2 \citep{kayama_spatially_2022,kayama_x-ray_2025}.
Therefore, we consider $B = $ 5--40 \uG\ in this paper, a range that encompasses all inferred values of $B$ in the large scale jets of SS~433.
Depending on the jet models \citep{hess_2024_ss43}, \vush\ could be 31,000--76,000~\kms, enabling to achieve a PeV-scale cutoff energy with $\eta_{\rm B} \geq 1$.
If we adopt \vush\ = 2,000--28,000 \kms, derived using the lower limit of \vd\ = 0.021$c$ \citep{hess_2024_ss43}, particle acceleration up to $E_0 \sim $ 200 TeV, as inferred from the gamma-ray spectra, becomes achievable with $\eta_{\rm B} \sim 1$, which indicates the most efficient acceleration case.
In \figref{fig:eta}, we show an illustration of the parameter space relevant to this discussion.



The sharp, edge-like structure observed at the innermost knots (Figures~\ref{fig:image} and \ref{fig:projection}) supports the DSA mechanism.
Alternatively, other acceleration mechanisms, such as modified non-linear DSA \citep[e.g., ][]{Caprioli2020}, turbulent acceleration, shear acceleration, and magnetic reconnection, are also plausible \citep{kaaret_x-ray_2024, safi-harb_hard_2022}.
Remarkably, recent polarization measurements by IXPE revealed that the magnetic field around the e0/head region is aligned with the jet axis \citep{kaaret_x-ray_2024}.
This likely indicates the presence of parallel shocks, as the shock front of e0 looks oriented perpendicular to the jet axis. 
This configuration is reminiscent of that observed in SNRs or powerful radio galaxies.

Here, we discuss the outer knots, e2, w2, and e3, of which the 3$\sigma$ upper limit speed was obtained as $v_{\rm app} <$ 0.012--0.019$c$.
The e2 and w2 knots are located 50--60 pc (0.5--0.6 deg for $d=5.5$ kpc) away from SS~433, and e3 is 100 pc (1 deg) away.
At these positions, the jet speed would be 0.045$c$ or 0.061$c$ in the constant-speed jet model, 
or the jet would likely decelerate with 0.014$c$, 0.0093$c$, 0.0052$c$ at e2, e3, and w2, respectively, in the decelerating and expanding jet model of \cite{hess_2024_ss43}, although the latter case seems unlikely in the e0/head and w1 knots.
Assuming these speeds as the upstream flow speeds (\vu) at e2 and w2 and using the measured \vapp\ (=\vsh), \vd\ can be derived as 
\begin{eqnarray}
v_{\rm d} = \frac{1}{4} \left( v_{\rm u} + 3 v_{\rm sh} \right), 
\end{eqnarray}
and \vush\ can be obtained by \eqref{eq:RH}.
We found \vush\ could reach 10,000--20,000 \kms\ for \vu\ = 0.045--0.061$c$ in the constant-velocity jet model (\tabref{tab:Vu}), suggesting that strong shock acceleration and/or reacceleration would be possible. 
In the decelerating jet model, we obtain \vush\ of 0--2,100~\kms\ at w2 and 660--5,500~\kms\ at e2.
This may indicate that particle acceleration at a strong shock is unlikely in the w2 knot, and the bright emission may be instead attributed to the locally enhanced magnetic field, as demonstrated in \cite{kayama_spatially_2022}.
Our constraint on e3, \vapp\ $<$ 0.019$c$, is consistent with that reported by \cite{sakemi_energy_2021}, indicating that the termination shock does not show significant motion.

\subsection{Jet dynamics and knot formation}

While the jet speed ($v_{\rm jet}$) is $0.26c$ at the launching site, we found the apparent velocity of the X-ray knots to be relatively small ($v_{\rm app} < $ 0.019--0.033$c$), located $\sim$30~pc away from the central compact binary. Given the detection of non-thermal X-ray and VHE gamma-ray emission at these knots \citep[e.g., ][]{abeysekara_very-high-energy_2018,hess_2024_ss43,safi-harb_hard_2022}, these observations suggest the existence of a quasi standing shock, likely formed by the jet recollimation process (i.e., the so-called recollimation shock).
The position of the recollimation shock is approximately determined by the balance between jet ram pressure and external gas pressure. Assuming an external pressure profile of $p_\mathrm{ext}(z)=p_e(z/z_e)^{-\eta}$, the recollimation shock is expected to occur at 
\begin{equation}
z_\mathrm{rec}\approx \left[\frac{(2-\eta)^2\xi_1}{4} \frac{L_j v_\mathrm{jet}}{\pi c^2 p_e z_e^\eta}\right]^{1/(2-\eta)} ,
\label{eq:recollimation}
\end{equation}
where $p_e$, $\eta$, $\xi_1$, and $L_j$ are 
pressure at a fiducial height $z_e$, a slope of the pressure profile, a constant ($\xi_1 \approx$ 0.7--1)\footnote{$\xi_1$ is defined as $(1- \hat{n} \cdot \beta_+ / \hat{n} \cdot \beta_-) \lesssim 1$, where $\beta_-$ and $\beta_+$ are the local fluid 3-velocity upstream and downstream of the shock, respectively, and $\hat{n}$ denotes the normal to the shock surface \citep{Bromberg2009ApJ...699.1274B}.}, and the jet luminosity, respectively \citep{sedov_similarity_1959,Komissarov1998MNRAS.297.1087K, Bromberg2009ApJ...699.1274B}.
It should be noted that this model is based on conical (non-precessing) jets. However, \cite{eichler_focusing_1983} showed that even a precessing jet, such as SS~433, can be recollimated by the ambient medium and focused along the jet axis at large scales \citep[see also, e.g., ][]{millas_relativistic_2019,ohmura_continuous_2021}, 
and thus we adopted the non-precessing jet model.
Here, we assume that the stellar wind from the companion star provides the external pressure as $p_{\rm ext} \sim 2\times 10^{-12} (z/1~{\rm pc})^{-\eta} ~ {\rm erg}~{\rm cm}^{-3}$ \citep{safi-harb_rosat_1997,panferov_jets_2017}
with $\eta = 0.6$ \citep{Bromberg2009ApJ...699.1274B}.
With $L_j \sim 10^{39} ~{\rm erg}~{\rm s}^{-1}$ \citep{dolan_ss_1997,fabrika_jets_2004},
we obtained $z_{\rm rec} \sim $ 20--30 pc using \eqref{eq:recollimation}. 
This is roughly consistent with the observed locations of the innermost knots
and the calculation in \cite{bowler_w50_2020}.
Therefore, the knots could be produced by the recollimation shock, which is reconciled with our measurement indicating the presence of a standing shock with $v_{\rm app}<$ 0.019--0.033$c$.

Based on magnetohydrodynamics (MHD) simulations dedicated for the SS~433/W50 system \citep{ohmura_continuous_2021}, 
it is indeed likely that recollimation shocks exist at the positions of the innermost knots. 
The flow launched from the microquasar should be decelerated down to $\lesssim$ 30\% of the initial jet speed, which yields $\lesssim$ 0.08$c$ at the innermost knot regions.
This is roughly consistent with our measurements of $v_{\rm u} \sim $ 0.04--0.15$c$ in the innermost knots, adopting \vd\ of 0.021$c$ and 0.045$c$ (\tabref{tab:Vu}).

The formation mechanism of the innermost knots remains elusive.
Although a standing recollimation shock provides a plausible explanation consistent with the observations, the formation of such shocks would be challenging in a precessing jet like that of SS~433, as demonstrated by \cite{monceau-baroux_relativistic_2014}.
In addition to the recollimation shock scenario, alternative mechanisms have been proposed, including
interactions between the jet and the SNR shell or molecular clouds
and
internal shock waves generated by collisions between jet components with different velocities \citep[e.g., ][]{yamamoto_physical_2022,sakemi_molecular_2023}.
Additionally, spine-sheath models, as proposed for AGN jets \citep[e.g., ][]{matsumoto_magnetic_2021}, may also be a possibility.
However, 
while jet models with a noticeable gradient of the bulk speed cannot be excluded, they likely require very modest plasma viscosity in the direction of the speed gradient. Space plasma is typically characterized by noticeable viscosity along magnetic field \citep{braginskii_transport_1965}, thus such jet configurations are more natural on small scales. On large scales, the toroidal component of the magnetic field tends to dominate, which should reduce the speed gradient.
Furthermore, the current \chandra\ data do not reveal any evidence for such sheath structures.

Our proper motion measurements show a hint of outward motion of the knots.
Future observations with next-generation telescopes, such as Advanced X-ray Imaging Satellite (AXIS) which offers an angular resolution of 1.5 arcsec across its wide 24-arcmin-diameter FoV \citep{reynolds_overview_2023}, will enable more precise measurements.
By combining our \chandra\ observations in 2023--2024 (with a PSF of 0.5 arcsec) with an AXIS observation in 2035, the innermost knots with an expected motion of $\sim$ 0.2 arcsec~yr$^{-1}$ would shift by $\sim 2$ arcsec over the interval \citep{safi-harb_stellar_2023}.
This will allow for more stringent constraints on the upstream velocities of the shock, providing new insights into the nature of knot dynamics and particle acceleration mechanisms in the SS~433/W50 system.
Furthermore, investigating detailed spatial structures of jets and lobes in the Galactic microquasar, which is not feasible for most of extragalactic sources, will provide critical insights into the mechanisms of particle acceleration, propagation, and/or interactions with the surrounding materials of relativistic jets in active galactic nuclei.

\begin{acknowledgments}
We thank the anonymous referee for the helpful comments.
We are grateful to Takahiro Sudoh for his involvement in the proposal and early stages of the project.
We also thank Haruka Sakemi, Mami Machida, Susumu Inoue, Jiro Shimoda, and Shinya Yamada for fruitful discussions.
This research employs a list of Chandra datasets, obtained by the Chandra X-ray Observatory, contained in~\dataset[DOI:10.25574/cdc.444]{https://doi.org/10.25574/cdc.444}.
This work was supported by the Japan Society for the Promotion of Science (JSPS) KAKENHI grant Nos. JP22K14064 and JP24H01819 (NT) and
JP21H04493 (TGT). 
DK acknowledges support by RSF grant No. 24-12-00457.
KM acknowledges support from Chandra Cycle 23 and 24 GO programs (SAO GO3-23021X and GO3-24028X). 
SSH acknowledges support from the Natural Sciences and Engineering Research Council of Canada (NSERC) through the Canada Research Chairs and the Discovery Grants programs. DK and FA acknowledge support from the Sichuan Science and Technology Department (under grant number 2024JDHJ0001).
\end{acknowledgments}

\begin{contribution}


NT was responsible for the data analysis and writing and submitting the manuscript.
YI and DK were responsible for the interpretation and edited the manuscript.
KM and SSH managed the entire X-ray observation project of SS~433/W50 and edited and reviewed the manuscript.
SSH, LON, and BMI provided the formal analysis and validation. 
TT, BMI, KK, TGT, HU, TF, and FA contributed to drafting the observation proposals and the discussion of this manuscript.


\end{contribution}

%
\facilities{CXO, XMM-Newton}


\software{CIAO \citep[v4.14, ][]{fruscione_ciao_2006},
XMM-SAS (v.20.0) }



\appendix

\section{X-ray dataset}

\tabref{tab:dataset} summarizes the X-ray observation data utilized in this paper.

\begin{table}[ht!]
\begin{center}
\caption{Observation log}
\begin{tabular}{cccccccc}
\hline \hline 
    Target & Instrument & ObsID & Effective exposure & Date & RA  & Dec  & Roll  \\ 
        & & & (ks) & & (deg) & (deg) & (deg) \\ 
    \hline 
    %
\multicolumn{8}{c}{New \chandra\ data} \\ \hline
 w1 & \chandra\ ACIS-I & 25142 &  19.8 &  2023-07-19 &  287.69 &  5.04 &  202.35 \\    
 w1 & \chandra\ ACIS-I &  27957 &  19.6 &  2023-07-21 &  287.69 &  5.04 &  205.79 \\    
 w2 & \chandra\ ACIS-I &  25143 &  21.2 &  2023-10-02 &  287.44 &  5.07 &  272.19 \\    
 w2 & \chandra\ ACIS-I &  28943 &  10.3 &  2023-10-08 &  287.44 &  5.07 &  270.19 \\    
 w2 & \chandra\ ACIS-I &  28973 &  8.8 &  2023-11-08 &  287.44 &  5.07 &  292.19 \\    
 e0 & \chandra\ ACIS-I &  26533 &  19.7 &  2024-07-13 &  288.29 &  4.93 &  175.20 \\    
 e0 & \chandra\ ACIS-I &  29477 &  20.3 &  2024-07-13 &  288.29 &  4.93 &  175.20 \\    
 e0 & \chandra\ ACIS-I &  27064 &  29.9 &  2024-09-12 &  288.29 &  4.93 &  263.27 \\    
 e2 & \chandra\ ACIS-I &  26534 &  13.6 &  2024-08-23 &  288.53 &  4.93 &  249.42 \\    
 e2 & \chandra\ ACIS-I &  29510 &  35.3 &  2024-08-25 &  288.53 &  4.93 &  251.37 \\  
    %
    \hline 
    \multicolumn{8}{c}{Archival \chandra\ data} \\ \hline
 SS 433 & \chandra\  ACIS-S &  3790$^\dagger$ &  58.1 &  2003-07-10 &  287.95 &  4.98 &  183.15  \\ 
 w2 & \chandra\ ACIS-I &  3843$^\ast$ &  71.2 &  2003-08-21 &  287.43 &  5.02 &  252.21 \\ 
 e3 & \chandra\ ACIS-I &  21225 &  11.4 &  2019-08-15 &  288.96 &  4.84 &  240.2 \\    
 e3 & \chandra\ ACIS-I &  22141 &  39.6 &  2019-08-10 &  288.96 &  4.84 &  240.21 \\    
 e3 & \chandra\ ACIS-I &  22683 &  44.5 &  2019-08-16 &  288.96 &  4.84 &  240.21 \\    
    \hline 
    %
    %
    \multicolumn{8}{c}{Archival \xmm\ data} \\ \hline
    e2 & XMM MOS \& PN & 0075140401 &  32.5 &  2004-09-30 &  288.55 &  4.93 & --- \\    
    e3 &  XMM MOS \& PN & 0075140501 &  31.3 &  2004-10-04 &  288.98 &  4.86  & --- \\
    \hline
    \end{tabular}
    \label{tab:dataset}
\end{center}
    \tablecomments{ \\
    $\dagger$ Partially covering e0 with the off-axis angle of 16\arcmin.
    \\
    $\ast$ Partially covering w1 with the off-axis angle of 14\arcmin.
}
\end{table}

\section{Velocity estimation}

\begin{figure}[b!]
\plottwo{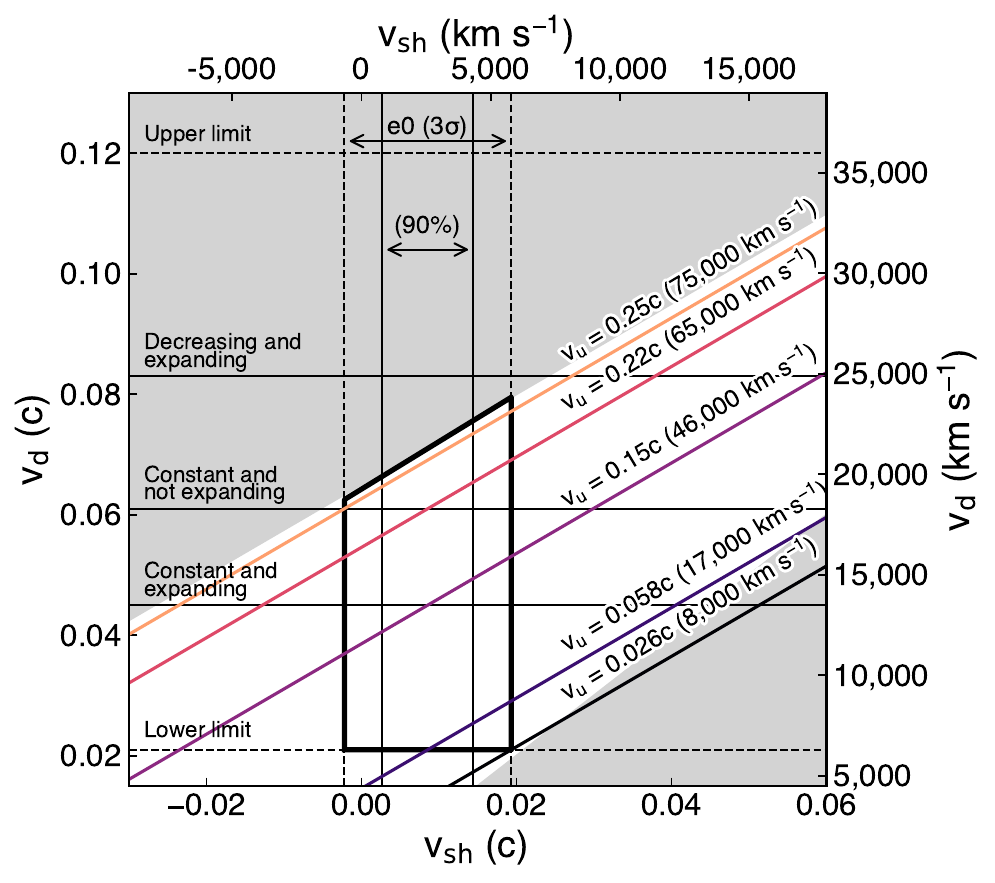}{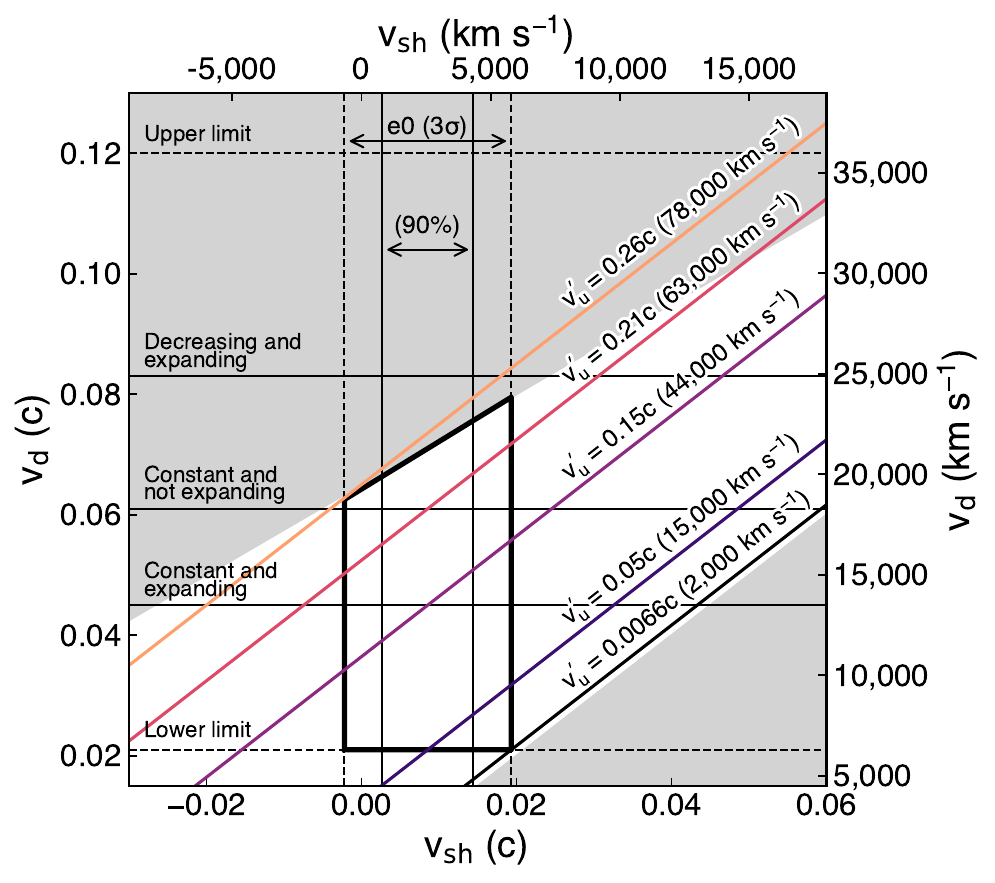}
\plottwo{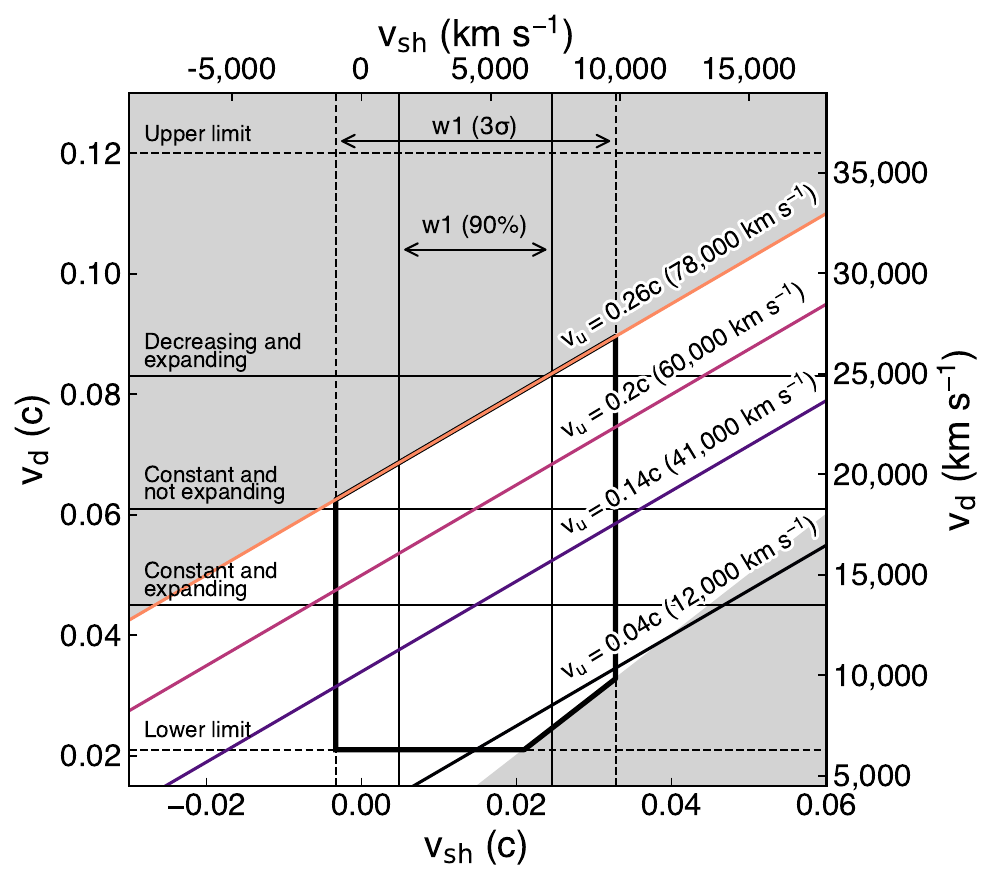}{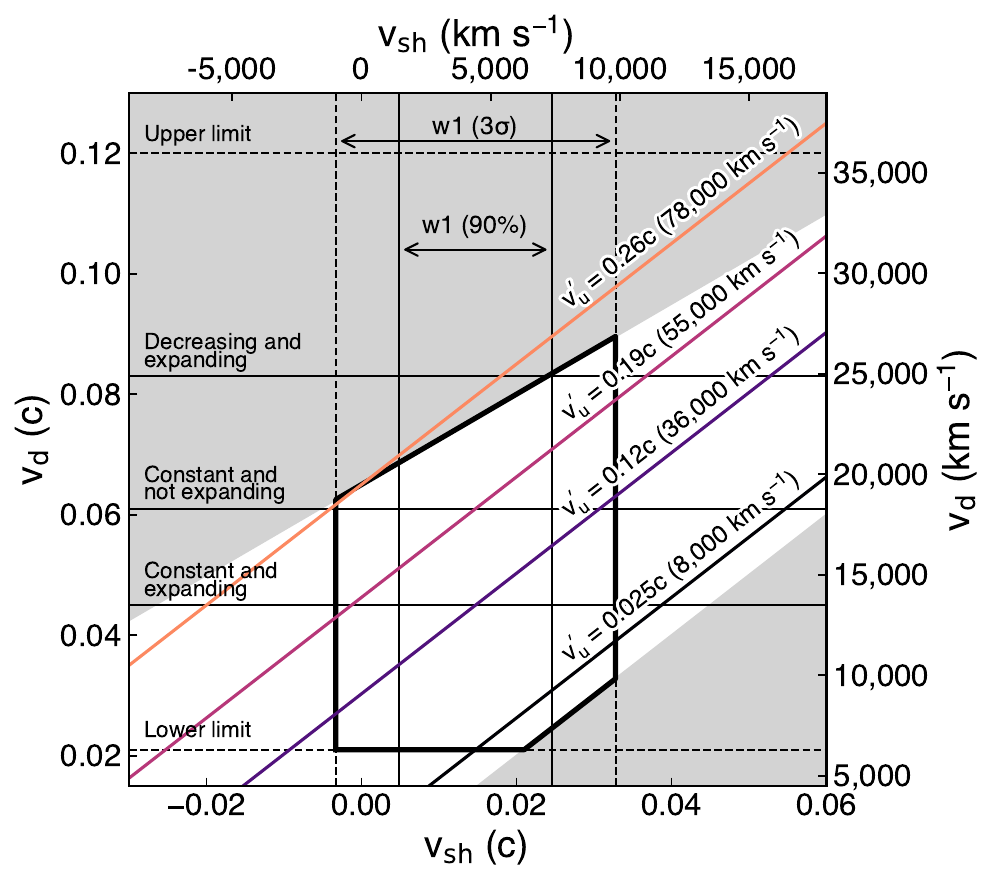}
\caption{
The constraints on the upstream velocity at the observer frame \vu\ (left) and at the shock rest frame \vush\ (right) as functions of $v_{\rm sh}$ and $v_{\rm d}$
(top: e0 and bottom: w1).
The vertical solid and dashed lines show the 90\% and 3$\sigma$ uncertainties of \vsh, respectively.
The horizontal lines show the constraints on the downstream velocity, assuming decreasing speed profile and expanding jet (\vd\ = 0.083$c$), constant speed and non-expanding jet (0.061$c$), and constant speed and expanding jet (0.045$c$), and the lower and upper limits including the systematic uncertainties (0.021$c$ and 0.12$c$) 
in \cite{hess_2024_ss43}.
The region enclosed by the tick black line indicates the allowed parameter space.
\label{fig:Vu}}
\end{figure}

\figref{fig:Vu} presents \vu\ and \vush\ calculated by Equations~\ref{eq:vu} and \ref{eq:RH}, respectively, and overlaid with the constraints of $v_{\rm sh}$ (this work) and \vd\ \citep{hess_2024_ss43}.
As the jet speed at the launching site of the microquasar SS~433 is measured as $\sim 0.26 c$ \citep[e.g., ][]{marshall_high-resolution_2002},
$ 0 \leq v_{\rm u} \leq 0.26c$ and $0 \leq v^\prime_{\rm u} \leq 0.26c$ are obvious and thus masked in \figref{fig:Vu}.
\tabref{tab:Vu} summarizes the obtained velocities.

\begin{table}[ht!]
\begin{center}
\caption{Velocity measurements of the nonthermal X-ray knots}
\begin{tabular}{ c | ccc |  ccc | ccc }
\hline \hline 
\multirow{2}{*}{Knot}  & \multicolumn{3}{c |}{\vd ($c$) } &  \multicolumn{3}{c | }{\vu ($c$)} &  \multicolumn{3}{ c  }{\vush\ ($c$)} \\ %
 &   Best fit & 90\% C.L. & 3$\sigma$ C.L. &  Best fit & 90\% C.L. & 3$\sigma$ C.L. &  Best fit & 90\% C.L. & 3$\sigma$ C.L. \\
\hline
\multirow{3}{*}{e0}  
& 0.021$^{\rm (a)}$ & --- & --- & 0.058 & $\pm$ 0.018 & $\pm$  0.032   &   0.050 & $\pm$ 0.024 & $\pm$  0.043  \\ 
& 0.045$^{\rm (b)}$ & --- & --- & 0.15 & $\pm$ 0.02 & $\pm$  0.03   &   0.15 & $\pm$ 0.02 & $\pm$  0.04  \\ 
& 0.061$^{\rm (c)}$ & --- & --- & 0.22 & $\pm$ 0.02 & $\pm$  0.03   &   0.21 & $\pm$ 0.02 & $\pm$  0.04  \\ 
\hline
\multirow{4}{*}{w1}  
& 0.021$^{\rm (a)}$ & --- & --- & 0.040 & $\pm$ 0.030 & $\pm$  0.054   &   0.025 & $\pm$ 0.040 & $\pm$  0.072  \\ 
& 0.045$^{\rm (b)}$ & --- & --- & 0.14 & $\pm$ 0.03 & $\pm$  0.05   &   0.12 & $\pm$ 0.04 & $\pm$  0.07  \\ 
& 0.061$^{\rm (c)}$ & --- & --- & 0.20 & $\pm$ 0.03 & $\pm$  0.05   &   0.19 & $\pm$ 0.04 & $\pm$  0.07  \\ 
& 0.083$^{\rm (d)}$ & --- & --- & 0.29 & $\pm$ 0.03 & $\pm$  0.05   &   0.27 & $\pm$ 0.04 & $\pm$  0.07  \\ 
\hline 
\multirow{3}{*}{e2} 
& 0.0063 & $\pm$ 0.0033 & $\pm$  0.0060  & 0.014$^{\rm (d)}$  & --- & --- &   0.010 & $\pm$ 0.017 & $\pm$  0.032  \\ 
& 0.014 & $\pm$ 0.003 & $\pm$  0.006  & 0.045$^{\rm (b)}$  & --- & --- &   0.041 & $\pm$ 0.017 & $\pm$  0.032  \\ 
& 0.018 & $\pm$ 0.003 & $\pm$  0.006  & 0.061$^{\rm (c)}$  & --- & --- &   0.057 & $\pm$ 0.017 & $\pm$  0.032  \\ 
\hline 
\multirow{3}{*}{w2}
& -0.0024 & $\pm$ 0.0051 & $\pm$  0.0093  & 0.0052$^{\rm (d)}$  & --- & --- &   0.010 & $\pm$ 0.027 & $\pm$  0.050  \\ 
& 0.0076 & $\pm$ 0.0051 & $\pm$  0.0093  & 0.045$^{\rm (b)}$  & --- & --- &   0.050 & $\pm$ 0.027 & $\pm$  0.050  \\ 
& 0.012 & $\pm$ 0.005 & $\pm$  0.009  & 0.061$^{\rm (c)}$  & --- & --- &   0.066 & $\pm$ 0.027 & $\pm$  0.050  \\ 
\hline
    \end{tabular}
    \label{tab:Vu}
    \end{center}
    \tablecomments{
    $d=5.5$~kpc is assumed. 
    In e0 and w1, the downstream velocity is assumed to be (a) the lower limit including the systematic uncertainty, 
    (b) the one derived in the constant speed and expanding jet model,
    (c) the constant speed and non-expanding jet, 
    and (d) the decreasing speed and expanding jet \citep{hess_2024_ss43}.
    This downstream speed corresponds to the upstream value in the e2 and w2 knots in cases (b) and (c), while the decreased velocity is derived in the decelerating jet model (d) (see also the text).
}
\end{table}
\bibliography{references,references2}
\bibliographystyle{aasjournalv7}



\end{document}